# Ballistic-like space-charge-limited currents in halide perovskites at room temperature

Osbel Almora[1,2,a], Daniel Miravet[3], Marisé García-Batlle[1] and Germà Garcia-Belmonte[1]

**AFFILIATIONS**

[1] Institute of Advanced Materials (INAM), Universitat Jaume I (UJI), 12006 Castelló, Spain

[2] Erlangen Graduate School of Advanced Optical Technologies (SAOT), Friedrich-Alexander Universität Erlangen-Nürnberg, 91052 Erlangen, Germany

[3] Quantum Theory Group, Department of Physics, University of Ottawa, Ottawa K1N 7N9, Canada

[a] Author to whom correspondence should be addressed: *almora@uji.es*,

## ABSTRACT

The emergence of halide perovskites in photovoltaics has diversified the research on this material family and extended their application towards several fields in the optoelectronics, such as photo- and ionizing-radiation-detectors. One of the most basic characterization protocols consist on measuring the dark current-voltage ($J - V$) curve of symmetrically contacted samples for identifying the different regimes of space-charge-limited current (SCLC). Customarily, $J \propto V^n$ indicate the Mott-Gurney law when $n \approx 2$, or the Child-Langmuir ballistic regime of SCLC when $n \sim 3/2$. The latter has been found in perovskite samples. Herein, we start by discussing the interpretation of $J \propto V^{3/2}$ in relation to the masking effect of the dual electronic-ionic conductivity in halide perovskites. However, we do not discard the actual occurrence of SCLC transport with ballistic-like trends. Therefore, we introduce the models of quasi-ballistic velocity-dependent dissipation (QvD) and the ballistic-like voltage-dependent mobility (BVM) regimes of SCLC. The QvD model is shown to better describe electronic kinetics, whereas the BVM model results suitable for describing both electronic or ionic kinetics in halide perovskites as a particular case of the Poole-Frenkel ionized-trap-assisted transport. The proposed formulations can be used as characterization of effective mobilities, charge carrier concentrations and times-of-flight from $J - V$ curves and resistance from impedance spectroscopy spectra.

**Introduction**

Halide perovskites, e.g. methylammonium lead triiodide (MAPbI$_3$), have emerged as one of the most attractive material families for optoelectronic applications, such as photovoltaics,[1] ionizing radiation detectors,[2] field-effect transistors,[3] memristors,[4] and energy storage.[5] Reason for that, is the optimal properties for photon absorption and charge transport, as well as the easy solution-based fabrication methods, which has motivated intensive research. At a material level, one of the most basic and commonly performed characterization techniques is the current density-voltage (*J-V*) measurement, which can inform on the transport mechanisms and properties. Notably, the study of the dark *J-V* curves of halide perovskites results particularly puzzling due to the dual ionic-electronic conductivity of these materials, resulting in many artefacts and anomalous behaviours, typically known as the *J-V* hysteresis effect.

The main transport mechanisms usually reported in symmetrically contacted samples (no built-in voltages $V_{bi}$ at the electrodes) of hybrid perovskites are the ohmic and the space-charge-limited current (SCLC). In the ohmic regime, the current and the electrostatic potential ($\varphi$) are linear with the applied voltage and the position ($x$), respectively (see FIG. 1(a,b)). Differently, in the SCLC regime the current may deviate from the linear behaviour due to significant modification in the charge density profile as the electric field increases with the externally applied voltage.

The most common form of SCLC is that of the mobility regime, where the current follows the Mott-Gurney law ($J \propto V^2$).[6, 7] Within this regime, one can often find the trap-filling ($J \propto V^{>2}$) sub-regime,[8] which allows the estimation of the concentration of deep level trap defects.[7] Subsequently, upon high enough injection of charge carriers, the velocity saturation regime ($J \propto V$),[9, 10] takes place. These regimes are well known and their identification in hybrid halide perovskites is generally not questioned with the classic distribution of FIG. 1(a).

Another fundamentally different SCLC regime is that of the ballistic transport, as the Child-Langmuir law,[11, 12] where

$$J = \frac{4\epsilon_0\epsilon_r}{9 L^2} \sqrt{\frac{2Q}{M}} V^{3/2}. \qquad (1)$$

Here $\epsilon_0$ is the vacuum permittivity; $\epsilon_r$, the dielectric constant; $L$, the distance between electrodes; and $Q$ and $M$ are the charge and the effective mass of the charge carriers, respectively. In physical terms, unlike the mobility regime, in the ballistic regime no scattering is considered (the definition of mobility is not justified) and the maximum drift velocity results

$$v_d = \sqrt{\frac{2Q}{M} V} \qquad (2)$$

as a consequence of the conservative transformation of electrostatic energy into kinetic energy. This model was initially thought for vacuum conditions, and among semiconductors it has been considered for either low temperature conditions,[13] short enough distance[14] or time scale (near-ballistic regime).[15] Notably, in practice, with typical measurement conditions ($V \leq 100$ V, room temperature $T \sim 300$ K, electrode active area A>0.01 cm$^2$, $\epsilon_r$=23[16]) the ballistic currents of MAPbI$_3$ samples would result significantly high for $L$<10 μm (e.g. $J$>10 A cm$^{-2}$ at 10 V), as presented in Figure S1 in the supplementary material. In addition, if present, a bulk mechanism such as the SCLC is commonly overlapped by interface phenomena in thin film samples where the distance between electrodes is in the order of the diffusion and/or Debye lengths. Therefore, it makes sense to discard Child-Langmuir's ballistic SCLC transport as a major mechanism for operational currents in thin film samples.

In thick halide perovskite samples the SCLC regimes with trends $J \propto V^n$ and $1 < n < 2$ have been identified[17] which may suggest the presence of some sort of ballistic-like transport. For example, FIG. 1(c) shows three different perovskite samples showing some section of the dark *J-V* curve with $n \approx 3/2$, characterized in our previous and simultaneous works and measured in room conditions with a Keithley Source-meter 2612B.[18, 19] The chromium contacted 1-mm-thick polycrystalline (pc) pellet (Cr/pc-MAPbI$_3$/Cr)[18] reports the lower currents and the 2 mm-thick single crystal (sc) of CH$_3$NH$_3$PbBr$_3$ (Cr/sc-MAPbBr$_3$/Cr)[20] sample presents the shorter voltage section with the $J \propto V^{3/2}$ trend, whereas the platinum contacted 3 mm-thick CsPbBr$_3$ sample (Pt/sc-CsPbBr$_3$/Pt)[19] seems to behave the closest to the seemly ballistic comportment.

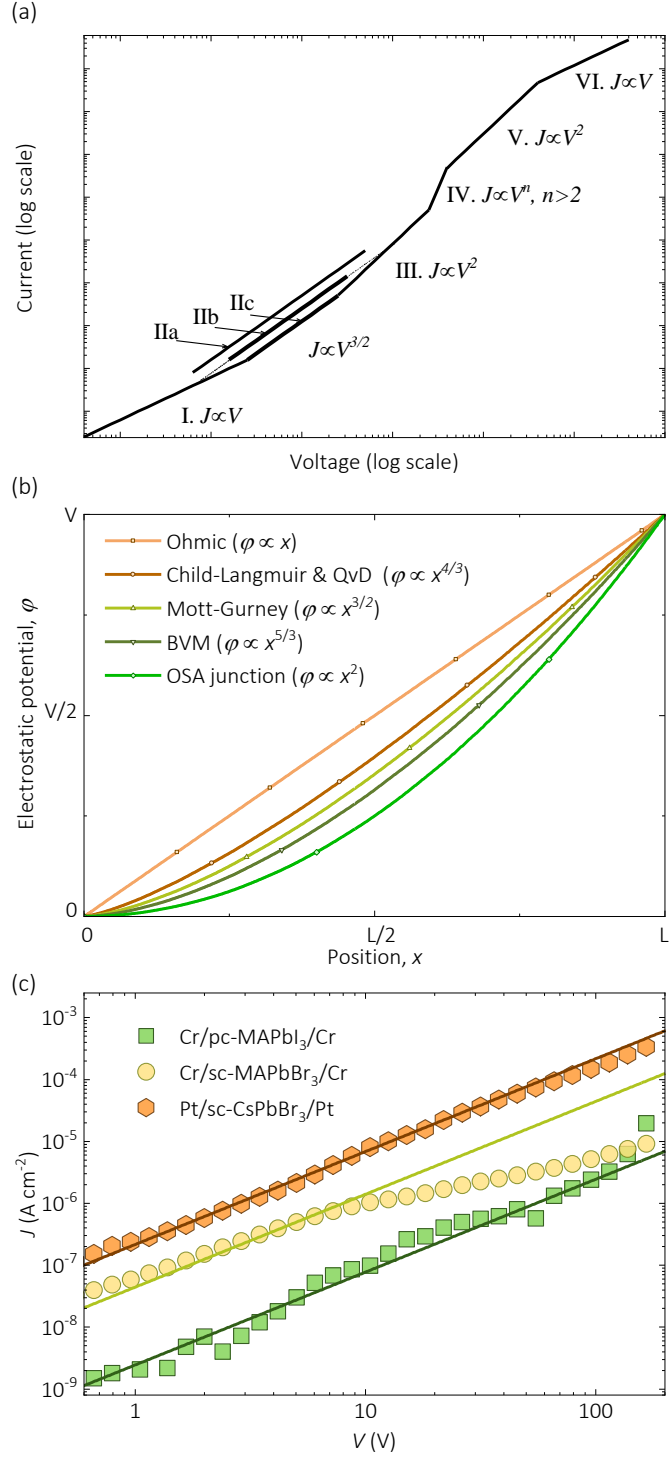

FIG. 1. Current density as a function of the external applied voltage. (a) Theoretical transport regimes: (I) ohmic, (IIa) ballistic Child-Langmuir law[11, 12] of SCLC, (IIb) quasi-ballistic velocity-dependent dissipation (QvD) model of SCLC, (IIc) ballistic-like voltage-dependent mobility (BVM) model of SCLC, (III) Mott-Gurney law[7] trap-empty SCLC, (IV) trap filling,[7] (V) Mott-Gurney law trap-filled[7] and (VI) velocity saturation.[7] (b)

Electrostatic potential as a function of position between electrodes separated a distance *L* upon application of voltage *V* for several configurations from ohmic to one-sided abrupt (OSA) junction, as indicated. (c) Experimental curves from various perovskite samples (dots) reported in the supplementary materials of our simultaneous works,[19, 21] licensed under Creative Commons Attribution Non-commercial (CC-BY-NC) licenses, and allometric fitting (lines) to the ballistic-like behavior ($J \propto V^{3/2}$).

In this work, first, we question the identification of the ballistic regime of SCLC in hybrid halide perovskite and analyse possible misleading influences of the ionic-electronic character of the conductivity in these materials. Secondly, we provide analytical formulations in two models that reproduce a similar behaviour ($J \propto V^{3/2}$) despite not being strictly ballistic regimes: the quasi-ballistic velocity-dependent dissipation (QvD) and the ballistic-like voltage-dependent mobility (BVM) regime of SCLC.

For all SCLC regimes, any trend from a dark *J-V* curve of halide perovskite samples should be taken with cautions, because of the hysteresis issues. For instance, Duijnstee et al.[22, 23] recently suggested the use of temperature-dependent-pulsed-voltage-SCLC measurements as a validating technique for ensuring the identification of one or another transport regime. Also, Alvar et al.[24] showed how the frequency dependence of the permittivity of thin film perovskites, and their dependence on voltage scan rate and temperature, influence the analysis of the SCLC. Moreover, the temperature-modulated-SCLC spectroscopy study of Pospisil et al.[25] found multicomponent deep trap states in pure perovskite crystals, assumingly caused by the formation of nanodomains due to the presence of the mobile species in the perovskites. These and more studies suggest that any claim of SCLC from *J-V* curves should be double checked for hysteresis or ion migration related artefacts, before extracting parameters. For instance, the *J-V* curves of FIG. 1(c) were not reproducible when long periods of pre-biasing preceded the voltage sweeps.

Regarding the classic ballistic regime of SCLC in halide perovskites, even if a $J \propto V^{3/2}$ trend is validated as hysteresis-free, there is little likelihood for the mechanism to be present at room temperatures. On the one hand, the condition of negligible energy dissipation for charge carriers is arguably unrealistic for transport from one electrode to the other, regardless whether they are electronic or ionic charge carriers. Notably, even though there is evidence of ballistic transport lengths ~200 nm in MAPbI$_3$ thin films, it has been reported in a time scale of 10-300 fs.[26-28] On the other hand, a fair assumption would be for the $J \propto V^{3/2}$ tendency to more likely be an intermediate sub-regime of transition between the ohmic regime ($J \propto V$) and the mobility regime ($J \propto V^{\geq 2}$), or between this latter and ohmic-seemly velocity saturation regime ($J \propto V$). However, a $J \propto V^{3/2}$ behaviour could occur in a way that resembles the SCLC ballistic mechanism at room temperature. For this to happen in the SCLC formalism, two main scenarios can be considered. First, a quasi-ballistic SCLC transport can occur in the case of an energy dissipation proportional to the velocity, resembling that due to Stoke's drag dissipation forces. Second, a ballistic-like SCLC transport takes place due to a behaviour of $v_d \propto V^{1/2}$ not related at all with energy conservation (ballistic regime) but with a bias-dependent energy dissipation (mobility regime).

**The quasi-ballistic velocity-dependent dissipation regime**

The core of the deduction of the classic SCLC resides in neglecting the diffusion currents (div**J**=0) and assuming that the total current density is mostly the drift component with a given relation between $v_d$ and the electrostatic potential $\varphi$ or the absolute value of the electric field $|\xi| = |d\varphi/dx|$. Mathematically, for a sample with distance *L* between electrodes at an external voltage *V* and $v_d \propto \varphi^p$, one can always

find a solution of the Poisson equation with a position-independent current such as $J = K V^{p+1} L^{-2}$, where $K$ is a constant (see Equation S10 in Section S1 of the supplementary material). Thus, $v_d \propto \varphi^{1/2}$ results in the Child-Langmuir ($J \propto V^{3/2}$)[11, 12] law with a potential $\varphi \propto x^{4/3}$ (see FIG. 1).

Unlike the classic ballistic approach, where all the electrostatic energy is converted into kinetic energy, a dissipation term is included as a correction in the QvD model. The energy dissipated during the transport of the charge carrier from one electrode to the other is considered in the form of $W = M\nu L v_d$, similarly to a Stoke's drag dissipation mechanism. Here, the momentum relaxation rate as $\nu = \nu_0 (\varphi/V_0)^{1/2}$ is such that it agrees with Teitel and Wilkins' conditions for the typical $v_d$ time overshoot, as previously considered for near-ballistic transport in one-valley semiconductors.[29] The characteristic dissipation frequency $\nu_0$ is in the order of the black-body radiation and $\nu$ increases with the potential, and thus $V$, over the threshold value $V_0$, attaining a maximum at the biased electrode. Moreover, $V_0$ is also expected to relate to $L$ balancing the dissipation energy in a form that $W$ only depends on the position $x$ through $\varphi$ and $v_d$. Subsequently, the energy conservation results as

$$\frac{M}{2} v_d^2 + \sqrt{\beta M Q \varphi} v_d = Q\varphi \tag{3a}$$

where the dimensionless QvD coefficient is

$$\beta = \frac{M \nu_0 \mu_\beta}{Q} \tag{3b}$$

and the effective mobility is $\mu_\beta = \nu_0 L^2 / V_0$. Notably, the use of effective mobilities has been earlier introduced in quasi-ballistic transport in high electron mobility transistors.[30] In FIG. 2(a), the typical values for $\beta$ as function of $\nu_0$ and $\mu_\beta$ for a charge carrier with the elementary charge $q$ and the electron mass $m_e$. The exact solution of Equation (3) is only different to (2) by a factor $\left((2 + \beta)^{1/2} \pm \beta^{1/2}\right)$ which is $2^{1/2}$ in the limit of $\beta \to 0$. Note that the solution with the plus must be discarded because the dissipation cannot increase the velocity,

For $\beta \ll 2$ (e.g., low temperatures, low mobilities, and/or low effective mass charge carriers), the dissipation does not significantly affects the Child-Langmuir law[11, 12] as in Equations (1) and (2). Thus, the use of the ballistic SCLC model would be justified with typical current values as in FIG. 1. On the other hand, the case could be a large dissipation for $\beta \gg 2$ (e.g. high temperatures, high mobilities, and/or high effective mass charge carriers). In this situation the ballistic regime would no longer be present. In between these two limits, with $\beta \sim 2$ the QvD-SCLC would effectively modify the drift velocity as

$$v_d = \left(\sqrt{2+\beta} - \sqrt{\beta}\right)\sqrt{\frac{Q\varphi}{M}} \qquad (4)$$

Taking the current density for the charge carrier concentration $N$ as

$$J = Q N v_d \qquad (5)$$

one can substitute (4) and (5) in the Poisson equation[11, 12]

$$\frac{d^2\varphi}{dx^2} = \frac{J}{\epsilon_0 \epsilon_r \left(\sqrt{2+\beta} - \sqrt{\beta}\right)} \sqrt{\frac{M}{Q\varphi}} \qquad (6)$$

whose solution between $x = 0$ and $x = L$, with $\varphi(0) = 0$ and $\varphi(L) = V$,[11, 12] results in the QvD-SCLC current density

$$J = \frac{4\epsilon_0 \epsilon_r \left(\sqrt{2+\beta} - \sqrt{\beta}\right)}{9L^2} \sqrt{\frac{Q}{M}} V^{3/2}. \qquad (7)$$

The values from Equation (7) are illustrated in FIG. 2(b) for electrons in a 1 mm-thick sample of MAPbI$_3$. For $\beta < 0.1$ the current is nearly that of the Child-Langmuir law, whereas it decreases as $\beta$ increases.

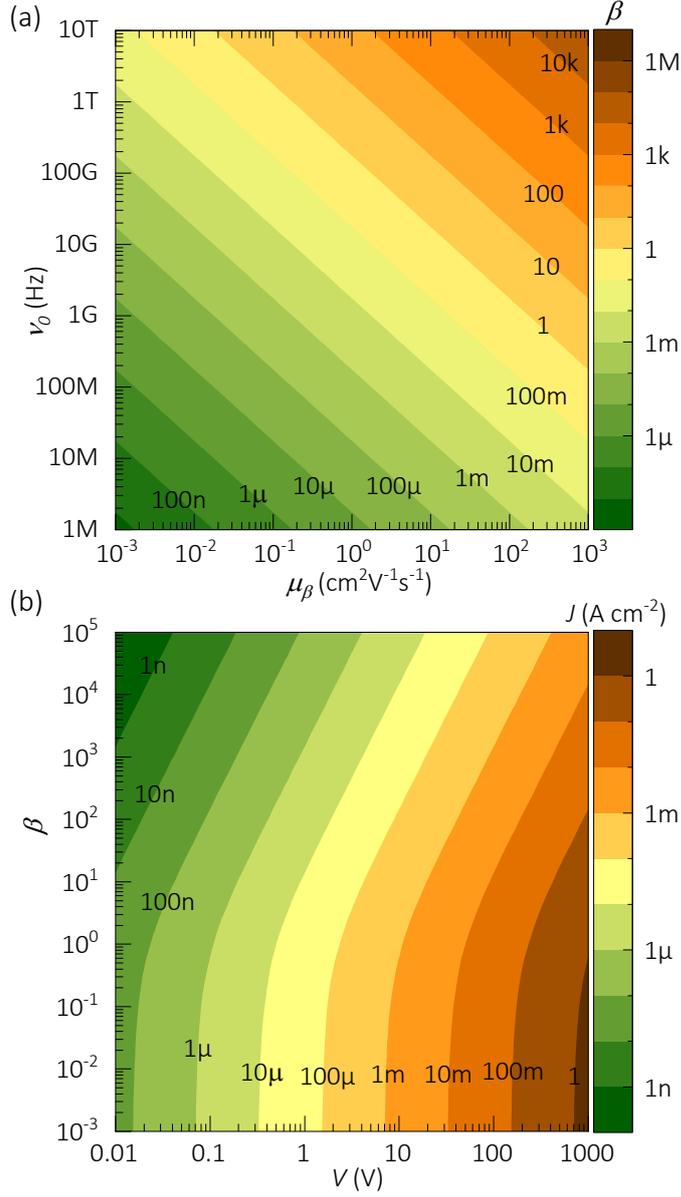

FIG. 2. The QvD-SCLCs approximation: (a) the QvD coefficient as in Equation (3b) and (b) the current density as in Equation (7) with $L$ =1 mm, $\epsilon_r$=23,[16] $M = m_e$ and $Q = q$.

In halide perovskites, the QvD-SCLC would be more appropriate for describing electronic kinetics due to their significantly higher mobility (~1 cm$^{-2}$V$^{-1}$s$^{-1}$), in comparison with that of the mobile ion charge carriers (<10$^{-6}$ cm$^{-2}$V$^{-1}$s$^{-1}$). The low mobility of the ions lowers $\beta$, which makes the currents to follow the Child-Langmuir law, that could attain unrealistically large ionic currents.

## The ballistic-like voltage-dependent mobility regime

The collapse of the QvD-SCLC model above presented occurs when the large dissipation ($\beta \gg 2$) produces current values out of the rage of those found in the experiment. In this case, the presence of SCLC with a $J \propto V^{3/2}$ trend could still be justified with a modified SCLC mobility regime. In the classic mobility regime, instead of the relation (2), the scattering is considered and the energy dissipation for the charge carriers is parameterized with the mobility $\mu$ as

$$v_d = \mu \xi \tag{8}$$

Similarly to the above reasoning, given a relation $v_d \propto (d\varphi/dx)^p$, one can always find a solution to the Poisson equation with a position-independent current such as $J = \varsigma V^{p+1} L^{-p-2}$, where $\varsigma$ is a constant (see Equation S23 in Section S2 of the supplementary material). Thus, while $v_d \propto d\varphi/dx$ results in the Mott-Gurney law ($J \propto V^2$)[6] with a potential $\varphi \propto x^{3/2}$, a dependency such as $v_d \propto (d\varphi/dx)^{1/2}$ implies a ballistic-like current ($J \propto V^{3/2}$) with a potential $\varphi \propto x^{5/3}$ (see FIG. 1).

For $v_d$ to be other than linear with $\xi$, the mobility should be field- and thus voltage-dependent, which is in agreement with earlier experimental reports for halide perovskites.[19, 31] Importantly, if $v_d$ is no longer linear with $\xi$, the definition of mobility is no longer satisfied. Nevertheless, one can propose an effective threshold mobility $\mu_0$ for the transition from the ohmic to the BVM regime of SCLC above an onset voltage $V_0$ so

$$\mu = \mu_0 \sqrt{\frac{V_0}{L\xi}}. \tag{9}$$

Subsequently, one can substitute Equation (9) in (8) to obtain the BVM drift velocity

$$v_d = \mu_0 \sqrt{\frac{V_0}{L} \frac{d\varphi}{dx}}. \tag{10}$$

The conceptual ideas behind the expressions (9) and (10) are presented in Section S2.1 of the supplementary material under two main assumptions: (i) the larger $L_i$ the larger $\mu$, where $L_i$ is Frenkel's equation[32] for the distance between the ions and their local potential maxima upon application of an external field; and (ii) the smaller $L_D$ the larger $\mu$, where $L_D$ is a Debye length for the accumulation of mobile ions towards the electrodes. In addition, Equation (10) can be deduced as a particular case of Poole-Frenkel's[9, 32, 33] ionized-trap-assisted transport where the charge carrier concentration is proportional to the field in a narrow bias range (see also Section S2.1). Summing up, it is the dual ionic-electronic conductivity of perovskites that enables the BVM regime of SCLC. The introduction of ionic space charges and mechanisms of field-dependent ionization is also suggested from the potential distribution, as presented in FIG. 1(b). The BVM model corresponds to an electrostatic potential situation somehow in between the Mott-Gurney law of SCLC (electronic bulk effect) and the quadratic case of the rectifying junctions (mobile ions depletion effect towards interfaces). Moreover, the idea of gradients in the transport properties such as in (9) has also been proposed in the ionic dynamic doping model,[18, 20, 34] where the biasing produce accumulation of mobile ions towards one electrode with slow kinetics. Consequently, one can substitute Equation (10) in (5) to solve the Poisson equation

$$\frac{d^2\varphi}{dx^2} = \frac{J}{\epsilon_0 \epsilon_r} \sqrt{\frac{L}{\mu_0^2 V_0 \frac{d\varphi}{dx}}} \qquad (11)$$

whose solution between $x = 0$ and $x = L$, with $\varphi(0) = 0$ and $\varphi(L) = V$,[11] results in the BVM current density

$$J = \frac{\epsilon_0 \epsilon_r \mu_0}{L^3} \sqrt{2V_0}\, V^{3/2} \qquad (12)$$

The current values for Equation (12) are presented for a 1 mm-thick MAPbI$_3$ sample with $V_0 = 10$ V in FIG. 3(a). There, one can see that the BVM model results in lower currents with respect to the mobility range, in comparison with the QvD-SCLC model and the Child-Langmuir law. This suggests that the BVM regime could be appropriate for describing both electronic and ionic kinetics in halide perovskite samples. In this same direction, the time-of-flight $\tau_{tof} = L/v_d$ can be obtained by substituting Equation (12) in (5) to obtain:

$$\tau_{tof} = \frac{L^4 Q N_\tau}{\epsilon_0 \epsilon_r \mu_0 \sqrt{2V_0}} V^{-3/2} \qquad (13)$$

where $N_\tau$ is an effective average charge carrier concentration. Typical $\tau_{tof}$ values are illustrated in FIG. 3(b) for a 1 mm-thick MAPbI$_3$ sample at 100 V. It can be seen that for single crystal samples, whose charge carrier concentration is typically lower than $10^{15}$ cm$^{-3}$, the electronic charge carriers ($\mu_e \sim 1$ cm$^{-2}$V$^{-1}$s$^{-1}$) and mobile ions ($\mu_i <10^{-5}$ cm$^{-2}$V$^{-1}$s$^{-1}$) would result in $\tau_{tof}$ values in the orders up to ms and ks, respectively. Purposely, a $\tau_{tof} \propto V^{-3/2}$ behaviour in the order of ks have been found in our simultaneous work on 2 mm-thick MAPbBr$_3$ single crystals with measurements of long term ionic-related current transients.[21]

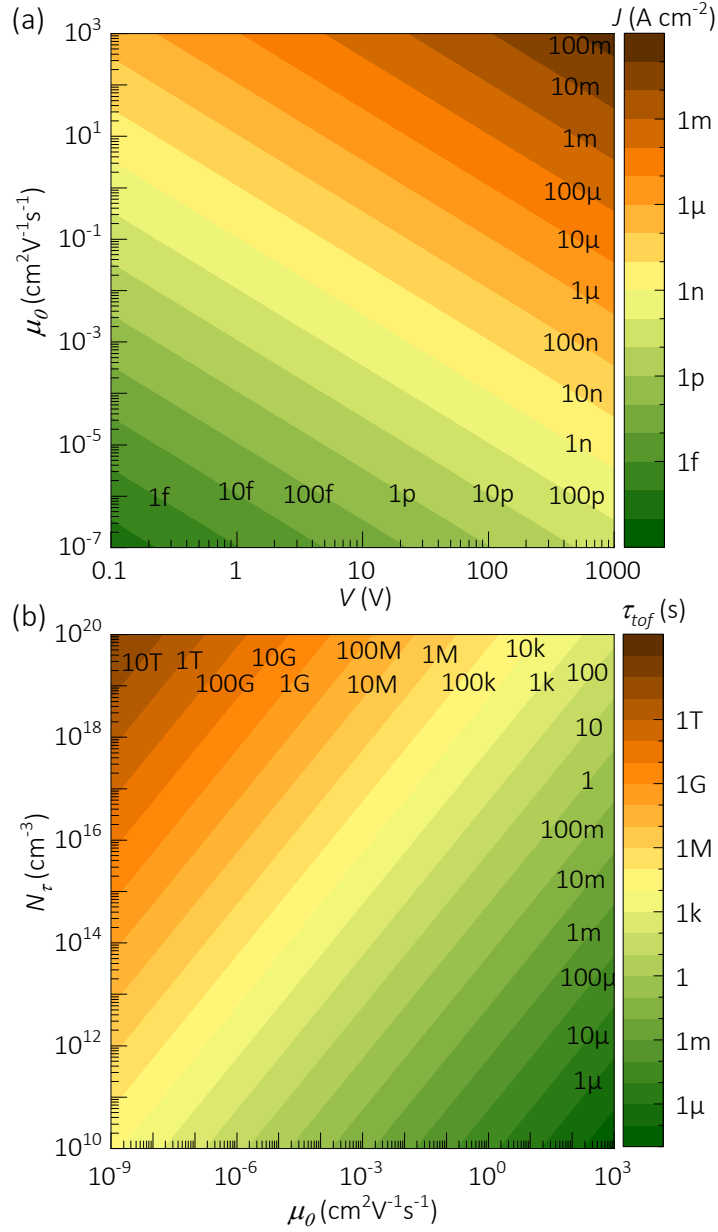

FIG. 3. The BVM regime of SCLC: (a) current density as in Equation (12) and (b) time-of-flight at 100 V as in Equation (13) with $L = 1$ mm, $\epsilon_r = 23$,[16] $V_0 = 10$ V.

The definition of differential resistance $R = (dJ/dV)^{-1}$ can also be applied to the current of Equation (12), resulting

$$R = \frac{\sqrt{2}L^3}{3\epsilon_0 \epsilon_r \mu_0 \sqrt{V_0}} V^{-1/2} \qquad (14)$$

Experimentally, Equation (14) relates to the resistance from impedance spectroscopy (IS) characterizations. Notably, the low frequency resistance from IS spectra have been found to behave as $R_{Lf} \propto V^{-1/2}$ in our simultaneous work on 3 mm-thick CsPbBr$_3$ single crystals.[19]

Importantly, discerning whether $J \propto V^{3/2}$ implies a QvD or a BVM regime of SCLC as Equations (7) and (12), respectively, can be problematic due to the experimental complications of measuring $v_0$ and finding $\beta$. However, an easier approach can be to compare the current density behaviour across a series of samples with different thicknesses. Accordingly, the linear trends $JL^2 \propto V^{3/2}$ and $JL^3 \propto V^{3/2}$ would be validating arguments for identifying QvD and BVM respectively.

## Conclusions

In summary, the use of the Child-Langmuir formalism of ballistic SCLC in halide perovskite has been reviewed and discussed. Our analyses suggest that, even though the reports on $J \propto V^{3/2}$ are most likely artefacts due to the so-called hysteresis effect in perovskite samples, $J \propto V^{3/2}$ behaviours should not be discarded as indicators of SCLC mechanisms in halide perovskites. In this regard, two models have been introduced. A quasi-ballistic regime of SCLC was analytically presented upon the assumption of an energy dissipation proportional to the drift velocity, similarly to the Stoke's drag mechanism. This model was found to most effectively describe electronic charge carrier kinetics in hybrid halide perovskites. On the other hand, we also introduced the analytical formalism for a ballistic-like mobility regime of SCLC based on a voltage dependency. By analysing the mobility dependency of the current in this latter model, we conclude that it could well describe electronic or ionic kinetics in hybrid halide perovskite. In addition, analogously to the classic Mott-Gurney law protocols, the introduced formalisms allow the extraction of key transport parameters from the $J - V$ curve and IS spectra, such as effective values for charge carrier mobility, concentration and time-of-flight.

See the supplementary material for the complete step-by-step deduction of the solutions of the Poisson equations and simulation.

We acknowledge the financial support from European Union's Horizon 2020 research and innovation program under the Photonics Public Private Partnership (www.photonics21.org) with the project PEROXIS under the grant agreement N° 871336. O.A. thanks Dr. Gebhard J. Matt for his feedback on the link to the Poole-Frenkel ionized-trap-assisted transport mechanism. M. G.-B. acknowledges Generalitat Valenciana for a grant (number GRISOLIAP/2018/073).

The data that support the findings of this study are available from the corresponding author upon reasonable request.

*Supporting Information:*

*Ballistic-like space-charge-limited currents in halide perovskites at room temperature*

Osbel Almora[1,2,a)], Daniel Miravet[3], Marisé García-Batlle[1] and Germà Garcia-Belmonte[1]

**AFFILIATIONS**

[1] Institute of Advanced Materials (INAM), Universitat Jaume I (UJI), 12006 Castelló, Spain

[2] Erlangen Graduate School of Advanced Optical Technologies (SAOT), Friedrich-Alexander Universität Erlangen-Nürnberg, 91052 Erlangen, Germany

[3] Quantum Theory Group, Department of Physics, University of Ottawa, Ottawa K1N 7N9, Canada

[a)] Author to whom correspondence should be addressed: *almora@uji.es*,

## S1. Potential-dependent drift velocity

Assuming $\varphi$ is the electrostatic potential of a sample of thickness $L$, where $\varphi(0) = 0$ and $\varphi(L) = V$, being $V$ the external applied voltage, then the drift velocity can be taken as

$$v_d = k\varphi^p \tag{S15}$$

where $p$ is a dimensionless constant and $k$ a constant whose units depend on $p$. Considering the current density $J$ independent of the position $x$ and $\varphi$, it can be expressed as

$$J = Q N v_d \tag{S16}$$

where $Q$ and $N$ are the charge and concentration of the charge carrier, respectively. By substituting Equation (S15) in (5) one obtains

$$N = \frac{J}{Qk} \varphi^{-p} \tag{S17}$$

Substituting Equation (S17) in the Poisson equation it results in

$$\frac{d^2\varphi}{dx^2} = \frac{K}{\varphi^p} \tag{S18}$$

where $K = J(k\epsilon_0\epsilon_r)^{-1}$. Then, we multiply Equation (S18) by $2d\varphi = 2(d\varphi/dx)dx$

$$2\frac{d\varphi}{dx}\left(\frac{d^2\varphi}{dx^2}dx\right) = 2\frac{K}{\varphi^p} d\varphi \tag{S19}$$

Subsequently, by integrating Equation (S19) we obtain

$$\left(\frac{d\varphi}{dx}\right)^2 - \left(\frac{d\varphi}{dx}\right)_0^2 = \frac{2K}{1-p} \varphi^{1-p} \tag{S20}$$

Neglecting the second term of the left member and elevating at the power of ½ Equation (S20), we obtain

$$\frac{d\varphi}{dx} = \left(\frac{2K}{1-p}\right)^{1/2} \varphi^{\frac{1-p}{2}} \tag{S21}$$

Equation (S21) can be reordered for integration between 0 and $x$, where $\varphi(0) = 0$ and $\varphi(x) = \varphi$, respectively, then

$$2\frac{\varphi^{\frac{p+1}{2}}}{(p+1)} = \left(\frac{2K}{1-p}\right)^{1/2} x \tag{S22}$$

Equation (S22) can be reordered and elevated at the power of $2(p+1)^{-1}$ to obtain the generic electrostatic potential as

$$\varphi = (p+1)^{\frac{2}{p+1}} \left(\frac{J}{2k\epsilon_0\epsilon_r(1-p)}\right)^{\frac{1}{p+1}} x^{\frac{2}{p+1}} \tag{S23}$$

Equation (S22) can also be reordered, elevated at the power of 2 and evaluated at $x = L$, where $\varphi(L) = V$, to obtain the generic current density

$$J = \kappa \frac{V^{p+1}}{L^2} \tag{S24}$$

where the constant $\kappa$ is

$$\kappa = 2k\epsilon_0\epsilon_r \frac{(1-p)}{(p+1)^2} \tag{S25}$$

In the ballistic regime, the Child-Langmuir law[1, 2] of space-charge-limited current (SCLC) uses $k = (2Q/M)^{1/2}$ and $p = 1/2$, then the electrostatic potential results as

$$\varphi = \left(\frac{3}{2}\right)^{\frac{4}{3}} \left(\frac{J}{\epsilon_0\epsilon_r}\sqrt{\frac{M}{2Q}}\right)^{\frac{2}{3}} x^{\frac{4}{3}} \tag{S26}$$

and the current density is

$$J = \frac{4\epsilon_0\epsilon_r\mu_0}{9L^2}\sqrt{\frac{2Q}{M}} V^{3/2} \tag{S27}$$

The ballistic current density as in Equation (S27) is illustrated in FIG. 1 for typical ranges of sample thickness and applied external voltage.

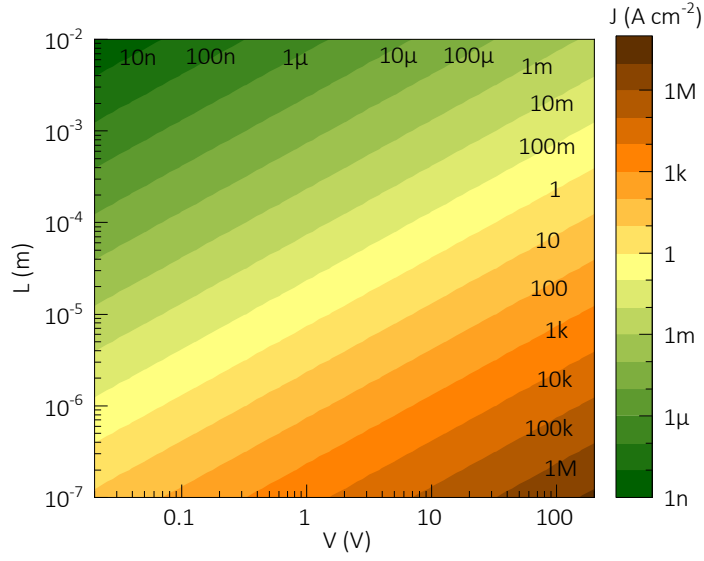

Figure S4: Current density as a function of the external applied voltage for the ballistic Child-Langmuir law[1, 2] of SCLC for a MAPbI$_3$ sample with $\epsilon_r$=23[3] and Q and M being the electron charge and mass, respectively, in Equation (S27).

In the quasi-ballistic velocity-dependent dissipation (QvD)[4] regime of SCLC, $k = (\sqrt{2+\beta} - \sqrt{\beta})\sqrt{\frac{Q}{M}}$ and $p = 1/2$, then the electrostatic potential results as

$$\varphi = \left(\frac{3}{2}\right)^{\frac{4}{3}} \left(\frac{J}{\epsilon_0 \epsilon_r} \sqrt{\frac{M}{Q}}\right)^{\frac{2}{3}} x^{\frac{4}{3}} \tag{S28}$$

and the current density is

$$J = \frac{4\epsilon_0 \epsilon_r \left(\sqrt{2+\beta} - \sqrt{\beta}\right)}{9L^2} \sqrt{\frac{Q}{M}} V^{3/2} \tag{S29}$$

## S2. Field-dependent drift velocity

Differently, the drift velocity can also be taken proportional to the absolute electric field $|\xi| = |d\varphi/dx|$, as

$$v_d = c \frac{d\varphi^p}{dx} \tag{S30}$$

where $c$ is a constant whose units depend on the dimensionless power $p$. Substituting Equation (S30) in (5) one obtains the total charge carrier concentration as:

$$N = \frac{J}{Qk} \frac{d\varphi^{-p}}{dx} \tag{S31}$$

Subsequently, substituting Equation (S31) in the Poisson equation it results in

$$\frac{d^2\varphi}{dx^2} = \frac{C}{\left(\frac{d\varphi}{dx}\right)^p} \tag{S32}$$

where $C = J(c\,\epsilon_0\epsilon_r)^{-1}$. Then, we multiply Equation (S32) by $\left(\frac{d\varphi}{dx}\right)^p dx$

$$\left(\frac{d\varphi}{dx}\right)^p \frac{d^2\varphi}{dx^2} dx = C\, dx \tag{S33}$$

Integrating Equation (S33) we obtain,

$$\left(\frac{d\varphi}{dx}\right)^{p+1} - \left(\frac{d\varphi}{dx}\right)_0^{p+1} = (p+1)Cx \tag{S34}$$

Neglecting the second term of the left member and elevating Equation (S34) at the power of $(p+1)^{-1}$ we obtain

$$\frac{d\varphi}{dx} = \left((p+1)Cx\right)^{\frac{1}{p+1}} \tag{S35}$$

Equation (S35) can be reordered for integration between $x = 0$ and $x$, where $\varphi(0) = 0$ and $\varphi(x) = \varphi$, respectively, then the electrostatic potential can be found as

$$\varphi = \frac{(p+1)^{\frac{p+2}{p+1}}}{(p+2)} \left(\frac{J}{c\,\epsilon_0\epsilon_r}\right)^{\frac{1}{p+1}} x^{\frac{p+2}{p+1}} \tag{S36}$$

Equation (S36) can be evaluated at $x = L$, where $\varphi(L) = V$, then the general current density results as

$$J = \varsigma \frac{V^{p+1}}{L^{p+2}} \tag{S37}$$

where the constant $\varsigma$ is

$$\varsigma = c\epsilon_0\epsilon_r \frac{(p+2)^{p+1}}{(p+1)^{p+2}} \tag{S38}$$

In the Mott-Gurney law[5] of SCLC, $c = \mu$ and $p = 1$, therefore, the electrostatic potential results as

$$\varphi = \frac{2}{3}\sqrt{\frac{2J}{\mu\,\epsilon_0\epsilon_r}}\,x^{\frac{3}{2}} \tag{S39}$$

and the current density is

$$J = \frac{9\epsilon_0\epsilon_r\mu}{8L^3}V^2 \tag{S40}$$

The current as in Equation (S40)(S27) is illustrated in Figure S5 for typical values.

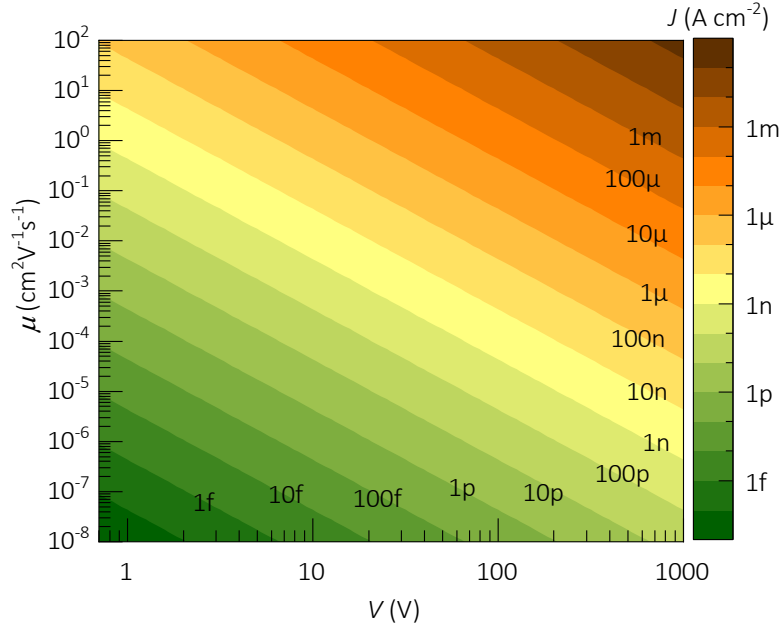

Figure S5: Current density as a function of the external applied voltage and mobility for the Mott-Gurney law[5] of mobility regime of SCLC for a MAPbI$_3$ sample with $\epsilon_r$=23[3] and L=1.0 mm in Equation (S40)(S27).

In the ballistic-like voltage-dependent mobility (BVM) regime, $c = \mu_0(V_0/L)^{1/2}$ and $p = 1/2$, then the electrostatic potential results as

$$\varphi = \frac{3}{5}\left(\frac{3}{2}\right)^{\frac{2}{3}}\left(\frac{J}{\epsilon_0\epsilon_r\mu_0}\sqrt{\frac{L}{V_0}}\right)^{\frac{2}{3}}x^{\frac{5}{3}} \tag{S41}$$

and, considering that $\sqrt{500/243} \approx \sqrt{2}$, the current density can be approximated to

$$J = \frac{\epsilon_0\epsilon_r\mu_0}{L^3}\sqrt{2V_0}V^{3/2} \tag{S42}$$

**S2.1. The onset voltage $V_0$ of the BVM regime of SCLC**

In the SCLC formalism, the $v_d \propto (d\varphi/dx)^{1/2}$ can explain a $J \propto V^{3/2}$, as above demonstrated. Typically, in the mobility regime the absolute value of the drift velocity is considered as

$$v_d = \mu \xi \tag{S43}$$

The use of Equation (S43) leads to the Mott-Gurney law.[5] However, assuming a transition from Ohmic to the BVM regime around an onset voltage $V_0$, the conjunction of both field-dependent ionization and accumulation of mobile ions towards the interface can be producing a voltage-dependent mobility as

$$\mu = \mu_0 \frac{L_i}{L_D} \tag{S44}$$

where $\mu_0$ is an effective mobility independent of field and position, $L_i$ is Frenkel's equation[6] for the distance between the ions and their local potential maxima upon application of an external field

$$L_i = \sqrt{\frac{Q}{\epsilon_0 \epsilon_r \xi}} \tag{S45}$$

and $L_D$ is the Debye length for the accumulation of mobile ions towards the electrodes

$$L_D = \sqrt{\frac{\epsilon_0 \epsilon_r k_B T}{Q^2 N_i}} \tag{S46}$$

In Equation (S46), $N_i$ is the mobile ions concentration, $k_B$ is the Boltzmann constant and $T$ is the temperature. By substituting Equations (S45) and (S46) in (S44), and multiplying by $(L/L)^{1/2}$ we obtain

$$\mu = \mu_0 \sqrt{\frac{V_0}{L\xi}} \tag{S47}$$

where the onset voltage comes after

$$V_0 = \frac{Q^3 N_i L}{\epsilon_0^2 \epsilon_r^2 k_B T} \tag{S48}$$

The values for $V_0$ are presented in Figure S6 for a MAPI sample at room temperature. Note that one may expect values in the range 1-10 V for a 1.0 mm thickness sample, meaning that the concentration of mobile ions towards the interface is around $10^{14}$ cm$^{-3}$.

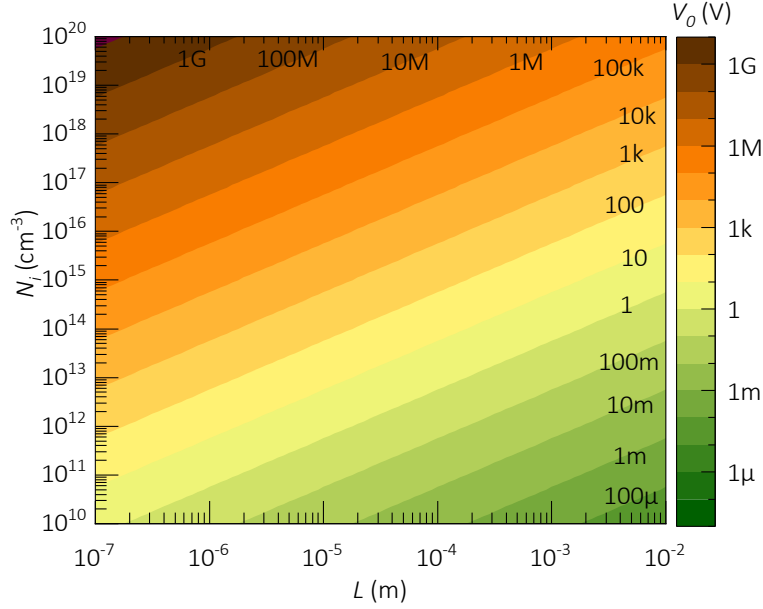

Figure S6: Onset voltage as a function of the distance between electrodes and the concentration of mobile ions towards the electrodes in the BVM regime of SCLC for a MAPbI$_3$ sample with $\epsilon_r$=23,[3] $T = 300$ K and $Q$ as the elementary charge in Equation (S48).

The BVM drift velocity can then be rewritten by substituting (S47) in (S43) as

$$v_d = \mu_0 \sqrt{\frac{V_0}{L}} \xi \tag{S49}$$

Alternatively, one can assume the BVM model as an approximation to a particular case of the Poole-Frenkel[6-8] ionized-trap-mediated transport when the field $\xi$ and the charge carrier profile $N$ meet certain specific criteria. For a start, we consider the Poole-Frenkel current density

$$J = \sigma_{PF} \exp\left[\frac{-q\phi}{k_B T}\right] \xi \exp\left[\sqrt{\frac{\xi}{\xi_{PF}}}\right] \tag{S50}$$

where $\sigma_{PF}$ is the Poole-Frenkel conductivity, $\phi$, the equilibrium potential barrier for the ionized traps and the Poole-Frenkel onset field is

$$\xi_{PF} = \frac{\pi\epsilon_0\epsilon_r k_B^2 T^2}{Q^3} \qquad (S51)$$

The electric field must fulfil two conditions for the BVM model to coincide with Equation (S50): *(i)* the field should be high enough that

$$\sqrt{\frac{\xi}{\xi_{PF}}} \gg 1 \qquad (S52)$$

in order to decrease the potential barrier in the Poole-Frenkel effect, but *(ii)* only in a narrow field range where the current is not critically exponential and the Equation (S50) can be approximated to the McLaurin expression as

$$J \cong \sigma_1 \exp\left[\frac{-q\phi}{k_B T}\right] \xi \left(1 + \sqrt{\frac{\xi}{\xi_{PF}}}\right) \qquad (S53)$$

where the effective BVM conductivity is taken as

$$\sigma_0 = \sigma_{PF} \exp\left[\frac{-q\phi}{k_B T}\right] \qquad (S54)$$

In addition, the charge carrier distribution should be approximately constant as

$$N = N_0 \sqrt{\frac{L}{\xi_{PF} V_0}} \qquad (S55)$$

where $V_0$ and $L$ have the same meanings as in Equation (S48) and $N_0$ is a threshold effective conductivity for the transition from ohmic to the BVM regime of SCLC. Subsequently, assuming (S52) and substituting Equation (S55) in (5) and (S53), the drift velocity can be approximated to Equation (S49) where the effective threshold mobility is assumed as

$$\mu_0 \cong \frac{\sigma_0}{N_0 Q} \qquad (S56)$$